\documentclass[sigconf]{acmart}

\usepackage{enumerate}
\usepackage{paralist}

\def\BibTeX{{\rm B\kern-.05em{\sc i\kern-.025em b}\kern-.08emT\kern-.1667em\lower.7ex\hbox{E}\kern-.125emX}}
    
\begin{document}

%
\title{CommentsRadar: Dive into Unique Data on All Comments on the Web}

%
\author{Sergey Nikolenko}

\orcid{0000-0001-7787-2251}
\affiliation{%
  \institution{Neuromation OU}
  \city{Tallinn}
  \country{Estonia}
  \postcode{10111}
}
\affiliation{%
  \institution{St. Petersburg Department of the Steklov Mathematical Institute}
  \city{St.Petersburg}
  \streetaddress{27 Fontanka street}
  \country{Russian Federation}
  \postcode{191023}
}
\email{sergey@logic.pdmi.ras.ru}

\author{Elena Tutubalina}
\affiliation{%
 \institution{Kazan Federal University}
 \streetaddress{18 Kremlyovskaya street}
 \city{Kazan}
 \state{}
 \postcode{420008}
 \country{Russian Federation}
 }
\affiliation{%
  \institution{St. Petersburg Department of the Steklov Mathematical Institute}
  \city{St.Petersburg}
  \streetaddress{27 Fontanka street}
  \country{Russian Federation}
  \postcode{191023}
}
\email{elvtutubalina@kpfu.ru}

\author{Zulfat Miftahutdinov}
\affiliation{%
 \institution{Kazan Federal University}
 \streetaddress{18 Kremlyovskaya street}
 \city{Kazan}
 \state{}
 \postcode{420008}
 \country{Russian Federation}
 }
\email{zulfatme@gmail.com}

\author{Eugene Beloded}
\affiliation{
\institution{SolidOpinion}
\country{United States}
}
\email{eugen.beloded@solidopinion.com}

%
\renewcommand{\shortauthors}{Nikolenko et al.}

%
\begin{abstract}
We introduce an entity-centric search engine \emph{CommentsRadar} that pairs entity queries with articles and user opinions covering a wide range of topics from top commented sites. The engine aggregates articles and comments for these articles, extracts named entities, links them together and with knowledge base entries, performs sentiment analysis, and aggregates the results, aiming to mine for temporal trends and other insights. In this work, we present the general engine, discuss the models used for all steps of this pipeline, and introduce several case studies that discover important insights from online commenting data.
\end{abstract}

%
%

\ccsdesc[500]{Computing methodologies~Natural language processing}
\ccsdesc[500]{Computing methodologies~Information extraction}
\ccsdesc[300]{Computing methodologies~Neural networks}
\ccsdesc[100]{Applied computing~IT architectures}

%
\keywords{opinion mining, user-generated texts, named entity recognition, sentiment analysis, web mining}

%

%
\maketitle

\section{Introduction}
The amount of text data being produced is overwhelming;
over 3 million blog posts are published on the Internet every day~\cite{stats2018}.
Hundreds of million people comment, reply to comments, and participate in online discussions. 
There is no question in the immense potential value of user comments, but retrieving and analyzing this valuable information presents a formidable challenge.

\begin{figure*}[t]\centering
\includegraphics[width=0.9\linewidth]{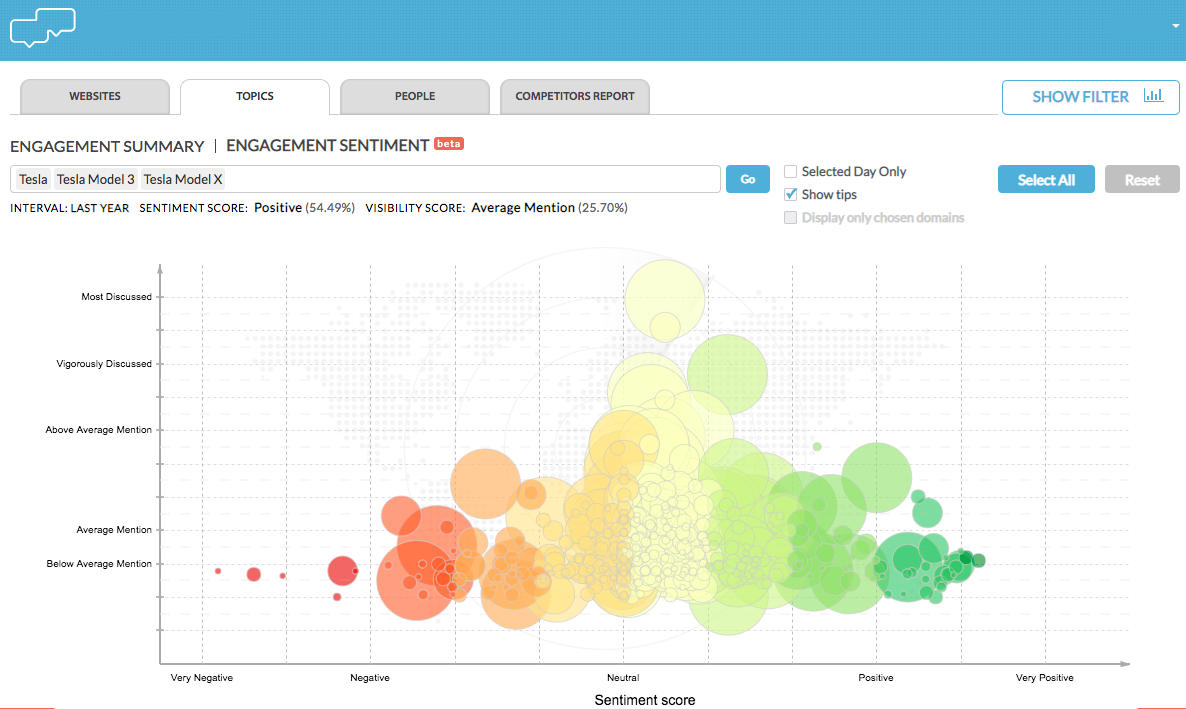}
\caption{Sentiment evaluation of comments about the \emph{Tesla} entity by the \emph{CommentsRadar} engine. Given an entity-centric query, the NLP engine retrieves articles and aggregate sentiment of user comments for them. Colored circles represent websites, their size corresponds to the number of articles and comments about this entity, and the horizontal axis shows sentiment.}
\label{fig:platform}
\end{figure*}

In this work, we present a comment search, user preference discovery, and recommendation engine \emph{CommentsRadar}~\cite{radar} that continuously scans the Internet and indexes all comments from select Web sites in order to identify trending topics and influencers.
The main goal is to provide information retrieval services above and beyond simple search over the comments. We would like to be able to find main topics of discussion, identify trends, extract entities that the texts discuss, link them between articles and comments, find out the general sentiment of user comments towards an entity or an article, find which topics a user is interested in, discover correlations between trending topics, advise online media which topics are likely to become popular, and so on.
For this purpose, \emph{CommentsRadar} combines modern NLP approaches based on deep neural networks to index and analyze the text of online articles along with user comments from top commenting Web sites. Articles and comments cover a wide range of topics, and we discover interesting correlations between different types of articles and entities, finding most influential users and sentiment of user opinions. As a result, our system can be and has been adopted for practical sentiment evaluation of user comments by property/publication over time; Fig.~\ref{fig:platform} shows a provisional interface of the system.

The paper is organized as follows. In Section~\ref{sec:models} we show the methods used in \emph{CommentsRadar}, including a general system pipeline (Section~\ref{sec:engine}), named entity recognition and linking models (Section~\ref{sec:ner}), and sentiment analysis (Section~\ref{sec:sent}). Section~\ref{sec:eval} presents our qualitative results in the form of three sample case studies made possible with the \emph{CommentsRadar} system, and Section~\ref{sec:concl} concludes the paper. 
\section{NLP Models in CommentsRadar}\label{sec:models}

\subsection{Engine Overview}\label{sec:engine}

\begin{figure}[t]\centering
\includegraphics[width=\linewidth]{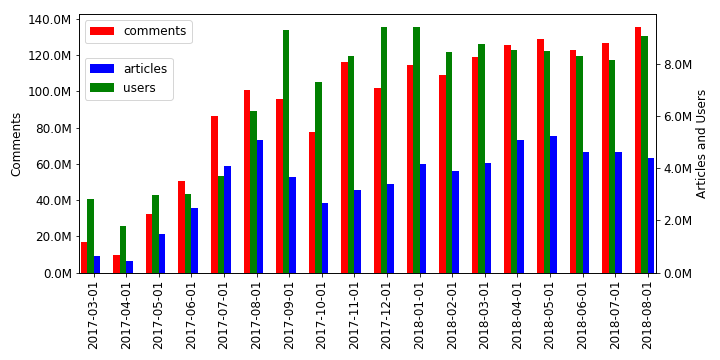}

\caption{Dataset statistics over time.}\label{fig:data}
\end{figure}

In this section, we present the pipeline of our engine named \emph{CommentsRadar} that automatically indexes and analyzes all comments written by users on the Web for a wide selection of websites. First, the system collects unstructured texts from top commenting sites in order to identify trending topics and influencers. The system obtains articles from different categories of websites, including:
\begin{enumerate}[(1)]
	\item ``News and Media'' sites, e.g., Fox News, Breitbart News, The Guardian;
    \item ``Government and Politics'' sites, e.g., The Daily Wire, The Hill, Wonkette;
    \item ``Arts and Entertainment'' sites, e.g., TMZ,  the Daily Express, The Avocado, and others.
\end{enumerate}
There are two basic strategies to aggregate online comments: querying social media such as \emph{Twitter} with entities extracted from articles or simply collecting all comments under a specific article. Many studies have used Twitter~\cite{agarwal2018geospatial,grvcar2017stance,karkulahti2016tracking,kwak2010twitter}, but we believe that the second strategy results in a less noisy dataset, aggregates more opinions, and makes it easier to connect opinions with sources. We crawl commenting sections of tens of thousands of Web sites and aggregate the comment texts. 
Supported commenting systems include Facebook, Disqus, Viafoura, Spot.IM, and others. All comments are linked to articles they relate to.
This pipeline produces a large dataset of user comments linked to the corresponding articles. Fig.~\ref{fig:data} shows the statistics over time; over 2017 and the first half of 2018 we have collected more than $1.67$B comments for more than $63.1$M articles. 

As for now, \emph{CommentsRadar} is in the final stages of the beta -- testing the system stability, speed capability, and the overall performance of the NLP pipeline described further. This product helps to deliver unique insight into the types of content consumed online, what online audiences care about. That's powerful information that can be leveraged by both Social Media Influencers, as well as the Agencies and Brand Advertisers that partner with them for greater ROI on their advertising campaigns with precision targeting.

The NLP problem here is to scan and categorize all this information. In the rest of this section, we describe in detail our NLP framework that:
\begin{inparaenum}[(1)]
\item extracts from the content of each article the named entities that describe parties, facts, events, and people involved in discussions;
\item performs entity linking between extracted entities and comments;
\item performs sentiment analysis on the comments;
\item aggregates and analyzes the results.
\end{inparaenum}
In the rest of this section, we describe the architectures used for each NLP component in detail.


\subsection{Named Entity Recognition and Linking}\label{sec:ner}

In order to select texts about a particular topic, we represent articles by named entities (NE) listing parties, facts, events, and people that come up in discussions. Entities are central to many web search queries; e.g.,~\citet{lin2012active} found that 57\% of queries have entities or entity categories and 28\% queries contain a reference to a website as an entity, while~\citet{guo2009named} demonstrated that 71\% of search queries contained named entities.

Named Entity Linking (NEL) is the task of assigning entity mentions in a text to corresponding entries in a knowledge base (KB). For example, the entity ``Barcelona'' in a sentence ``They have not lifted a European Trophy since 1991 when they beat Barcelona'' should refer to a knowledge base entity FC\_Barcelona (the football club) rather than the city. 
NEL is often regarded as crucial for natural language understanding and commonly used as preprocessing for tasks such as question answering~\cite{yih2015semantic} and semantic search~\cite{blanco2015fast}.

Given a text, the \emph{CommentsRadar} engine performs NER and NEL. For each article, first a NE recognizer extracts a set of entities: companies, people, products, buildings, location, date, and others. Then an entity linking model labels the entity mentions and provides unambiguous pointers to entities in a knowledge graph (KG) such as \emph{Wikipedia}.
For the first task, we examine several models, including
a production-ready pre-trained NER model \texttt{en\_core\_web\_md} from the \emph{spaCy} library~\cite{spacyfacts}
and our implementations of several LSTM-CRF models trained on OntoNotes~5.0~\cite{weischedel2013ontonotes}. OntoNotes 5.0 contains a variety of text domains including newswire, broadcast news, broadcast conversation, telephone conversation, and web data. The English newswire portion includes 300K of English \emph{Wall Street Journal} newswire. Following recent state-of-the-art models~\cite{lample2016neural,peters2017semi}, our NER models utilize pre-trained word embeddings, two bidirectional LSTM layers, and a conditional random field (CRF) loss~\cite{lafferty2001conditional}. We also experimented with ELMo (Embeddings from Language Models) representations derived from a bidirectional LSTM trained with a coupled language model (LM) objective~\cite{peters2018deep}.

The \emph{spaCy} model is based on context-sensitive embeddings and residual convolutional neural networks (CNN)~\cite{spacyfacts}. The NER model from the spaCy library was trained using multi-task learning. Aside from the NER task, the model learned on POS-tagging and dependency parsing tasks. 
It is reported to achieve F1 measure of $85.9\%$ on the OntoNotes corpus. We have attempted to reproduce this result on several subsets of OntoNotes 5.0 and seen some results differ from the reported.

Table~\ref{tab:spacy} presents the macro-averaged precision (P), recall (R), and F$_1$-measure (F) of the \emph{spaCy} model evaluated on various OntoNotes domains. \emph{SpaCy} shows high performance on Newswire and Broadcast news types but performs poorly on other types. Following recent state-of-the-art models~\cite{lample2016neural,peters2017semi}, we trained our own NER model with ELMo (Embeddings from Language Models) representations~\cite{peters2018deep}, bidirectional LSTM layers, and a conditional random field (CRF) loss~\cite{lafferty2001conditional}, getting 76.94\% F$_1$ on the Newswire subset of OntoNotes (see Table \ref{tab:ner_experiments}. 
The results are comparable, and Newswire and Broadcast news are exactly the type of text (and, more importantly, type of entities) we encounter most in \emph{CommentsRadar}, so in the system we used \emph{spaCy} since its license, unlike the OntoNotes corpus license, allows for commercial use.

\begin{table}[t]\centering
\caption{\emph{SpaCy} evaluation on the OntoNotes dataset.}
\label{tab:spacy}
\setlength{\tabcolsep}{6pt}
\begin{tabular}{|l|c|c|c|}
\hline
\textbf{Document type} & \textbf{Prec.} & \textbf{Rec.} & \textbf{F}  \\ \hline
Broadcast Conversations & 76.33 & 77.55 & 76.94 \\
Broadcast News & 87.90 & 88.46 & 88.18 \\
Magazine Genre & 76.92 & 81.08 & 78.95 \\
Newswire & 87.51 & 88.61 & 88.05 \\
Telephone conversations & 68.75 & 69.47 & 69.11 \\
Web Data & 74.47 & 73.79 & 73.63 \\
\hline
\end{tabular}
\end{table}

\begin{table}[t]
\caption{Experimental results on the Newswire subset of OntoNotes; 3-layer BiLSTM with char-level embeddings; other models, with GoogleNews embeddings and $512$ hidden units.}
\label{tab:ner_experiments}
\setlength{\tabcolsep}{5pt}
\begin{tabular}{|l|c|c|c|}
\hline
\textbf{Model} & \textbf{Prec.} & \textbf{Rec.} & \textbf{F}  \\ \hline
3-layer BiLSTM(+ELMo)-CRF &86.38 & 90.40 & 88.34 \\
1-layer BiLSTM-CRF & 83.88 & 86.92 & 85.37 \\
2-layer BiLSTM-CRF & 85.24 & 89.62 & 87.33 \\
2-layer BiGRU-CRF & 87.51 & 88.61 & 88.05 \\
\hline
\end{tabular}
\end{table}

For entity linking, we have compared two state of the art models:
the TAGME system~\cite{ferragina2010tagme} and a recently proposed multi-relational neural model~\cite{le2018improving}. TAGME is designed specifically for annotating ``on-the-fly'' short texts and queries with respect to Wikipedia. 
TAGME exploits the structure of the Wikipedia graph, scoring all possible relations between mentions and entities and then applying a voting scheme. 
We adopted the re-implementation of TAGME presented in~\cite{hasibi2016reproducibility}. Hasibi et al.~\cite{hasibi2016reproducibility} conclude that there are some technical challenges involved in the TAGME approach and some of the results did not reproduce even with the API provided by TAGME authors. \citet{le2018improving} used relations between entities as latent variables in a neural model, training it end-to-end and using CRF to assign the corresponding knowledge base entry to every mention. The authors made the code and pre-trained models publicly available. 

AIDA-CoNLL is manually annotated gold standard NEL dataset~\cite{hoffart2011robust} that contains news from Reuters Corpus V1 used for the CoNLL 2003 NER task. 
TAGME achieved 58.3\% micro-F1 measure on AIDA-CoNLL~\cite{cornolti2013framework}, and the multi-relational neural model achieved 93.07\%~\cite{le2018improving}. In our experiments with the \emph{Daily Mail} dataset (see Section~\ref{sec:dataset}), TagMe mapped 54.14\% of entity mentions to \emph{Wikipedia} pages, while the multi-relational neural model mapped 70.77\%. Thus, we have compared two models---state of the art from the ``pre-deep-learning'' era and current state of the art---and found considerable advantages of the latter, which we use in the system. Still, we believe that entity linking for \emph{CommentsRadar} can be further improved, and it is an important direction for further work.




\subsection{Sentiment Analysis}\label{sec:sent}

The goal of sentiment analysis is to identify and categorize the opinions or feelings expressed in a a piece of text, specifically to determine whether the writer's attitude in general or toward a specific topic is positive, negative, or neutral (as shown in the dashboard of \emph{CommentsRadar} on Fig.~\ref{fig:platform}).
A topic can be anything that users express opinions about: a celebrity (e.g., Taylor Swift), a policy (e.g., Obamacare), a product (e.g., Tesla Model S), an event (e.g., Formula 1 2018 VTB Russian Grand Prix), and so on. 
Numerous research studies develop sentiment models for a variety of domains and problem settings, ranging from very specific analysis of stance (e.g.,  whether the author of a text is in favor of or against a given target such as a political candidate \cite{mohammad2016semeval}) to general (e.g., SemEval 2016/2017 task 4~\cite{nakov2016semeval,rosenthal2017semeval}). Practical applications, however, are usually not as interested in the sentiment of a specific text 
as in averaged estimates of sentiment scores about mentioned entities in a set of articles and/or their comments over some time interval.

For \emph{CommentsRadar}, we have developed a sentiment model that classifies user opinions in general regardless of the topic. The neural network used for sentiment analysis consists of an embedding layer followed by CNNs with multiple filters of different lengths~\cite{kim2014convolutional}. To obtain local features from a text with CNNs, we used multiple filters of different lengths~\cite{kim2014convolutional}, replicating each filter on a hidden layer across the entire input vector, learning the same localized features in every part of the input and subsampling them, as usual for
one-dimensional CNNs, with max-over-time pooling layers that output the maximal value of a feature map over a time window. We used filters with window sizes $h\in \{3, 4, 5\}$ and $64$ feature maps each. Pooled features were fed to a fully connected layer with softmax activation. We also enhanced this model with pre-trained ELMo word representations \cite{peters2018deep}. 
%
We present an evaluation of our models on the SemEval 2017 Task 4 Subtask A dataset \cite{rosenthal2017semeval} which is publicly available; results are shown in Table~\ref{tab:sentexperiments}.
SemEval's primary measure was recall averaged across classes, and the DataStories model ranked first in Subtask A with average recall of $68.1$\%~\cite{baziotis2017datastories}. We see that our best model obtains average recall of $67.98$\% on the same dataset, and has the advantage of being generic and applicable to \emph{CommentsRadar} data. 

At the same time, there is still plenty of work left for on-going studies since user comments cover many different domains about various entities. Sentiment classification remains challenging: it is difficult to gather annotated training data for all of them. The current experiments are carried out on (i) our in-house annotated data about politicians and electronics, (ii) general 30M user-generated texts annotated with a distant supervision technique \cite{go2009twitter}, and (iii) publicly available datasets such as SemEval data, the Kaggle's toxic comment dataset \cite{jigsaw2018}, and the Yahoo news annotated comments corpus \cite{napoles2017finding}. The neural architecture utilizes the multi-domain framework \cite{ganin2016domain,liu2017adversarial,chen2018multinomial} to learn general features that are invariant across domains. Our extensive experiments have also shown that CNN is both efficient and effective over LSTM.
An important future extension would be to extract not only the overall sentiment of a post but also stance towards specific mentioned entities.


\begin{table}[t]
\centering
\renewcommand{\tabcolsep}{5pt}
\caption{Sentiment classification on SemEval 2017 Task 4 Subtask A; ELMO+CNN and CNN with 64 feature maps; CNN, BiGRU, and BiLSTM with GoogleNews embeddings.}
\begin{tabular}{|l|c|c|c|}
\hline
\textbf{Model} & \textbf{Prec.} & \textbf{Rec.} & \textbf{F}  \\ \hline
ELMO + CNN & 67.07 & 67.98 & 67.44 \\
CNN &  63.28 & 64.14 &  62.64 \\
BiGRU &  62.54 & 60.07 & 61.21 \\
BiLSTM & 61.3 & 59.75 & 60.47 \\
\hline
\end{tabular}
\label{tab:sentexperiments}
\end{table}

\section{Experimental Results}\label{sec:eval}

\label{sec:dataset}
In this section, we illustrate the operation of \emph{CommentsRadar} with both qualitative and quantitative results. 
We present two types of results: analyzing a website to figure out what readers like to discuss the most and find  influential commenters and studying the discussions about a particular event in order to understand user sentiment regarding it. 

We present three case studies: a middle-market tabloid newspaper (\emph{Daily Mail}\footnote{http://www.dailymail.co.uk}), an important political event (Brexit), and an Instagram celebrity (Kendall Jenner). \emph{Daily Mail}, a top selling newspaper with approximately $14.3$ million readers per month in the UK from October 2016 to September 2017~\cite{statista:2017}, covers a wide range of topics including politics, sports, celebrity news, science, and health stories. The United Kingdom EU membership referendum, also known as \emph{Brexit}, was held in the UK on June 23, 2016. British Prime Minister Theresa May signed an official letter invoking Article 50 on March 28, 2017 and thus making the UK's intention to leave the EU official. Finally, Kendall Jenner was one of the most popular figures on Instagram and also the subject of a media scandal during the period in question.

We explore \emph{Daily Mail} articles collected by \emph{CommentsRadar} from February 20 to August 8, 2017, linked with user comments from February 20 to June 20, 2017, with $29{,}101$ articles and $2{,}150{,}178$ comments in total, for a density (mean number of comments per article) of $74$, very high for an average over a large website. 
We ran our entire pipeline, including sentiment analysis, on Brexit-related posts published on the \emph{Daily Mail} website between February and June 2017.

\subsection{Daily Mail readership, topics, and influencers}

The primary goal of \emph{CommentsRadar} is to aggregate publisher- and user-generated content in order to identify trending topics and influencers. After named entity recognition and linking step, we have found that in the \emph{Daily Mail} dataset the average number of articles per entity was $3.2$, while the mean number of comments per entity was $477.33$.

Table~\ref{tab:entities} shows basic statistics: most commented entities sorted by the mean number of comments per article. The most engaging entities are related to the most common news sources in the scope of the data, e.g., Donald Trump. \emph{Instagram} is an anchor entity for many celebrity-related news. City and country names are obviously related to policial centers, and the \emph{Facebook} entity is very popular for two reasons: Facebook also appears as an anchor entity since many news originate there, and also \emph{Facebook Inc.} itself was a subject of political scandals during the spring and summer of 2017.

\begin{table}[t]
\centering
\setlength{\tabcolsep}{3pt}
\caption{Top entities w.r.t. mean number of comments per article (density).}
\label{tab:entities}
\begin{tabular}{|l|c|c|c|c|}
\hline
\textbf{Entity} & \textbf{\# comm.} & \textbf{\# art.} & \textbf{Density} & \textbf{Sentiment} \\
\hline
United Kingdom & 152,613 & 581 & 262.67 & -0.06 \\
Barack Obama & 123,019 & 474 & 259.54 & -0.22 \\
Donald Trump & 379,461 & 1,593 & 243.27 & -0.22 \\
Washington & 1,121,08 & 478 & 234.54 & -0.14 \\
White House & 191,989 & 836 & 229.65 & -0.20 \\
Facebook & 177,019 & 788 & 224.64 & -0.12 \\
United States & 215,350 & 1203 & 179.01 &  -0.10 \\
Russia & 120,042 & 634 & 189.34 & -0.15 \\
Europe & 102,889 & 601 & 171.96 & 0.02 \\
Manchester & 116,317 & 701 & 165.93 & 0.05 \\
London & 236,784 & 1,510 & 156.81 & 0.05 \\
New York & 100,434 & 845 & 118.85 & 0.003 \\
Instagram & 165,703 & 1,859 & 89.14 & 0.11 \\
\hline
\end{tabular}
\end{table}

\begin{table}[t]
\centering
\caption{Entities with highest positive or negative scores mentioned in more than average number of articles ($\ge 4$).}\vspace{-.2cm}
\label{tab:sententities}
\setlength{\tabcolsep}{2pt}
\begin{tabular}{|p{3.7cm}|c|r|c|}
\hline
\textbf{Entity} & \textbf{Sentiment} & \textbf{\# comm.} & \textbf{\# art.} \\ \hline
\multicolumn{4}{|c|}{Entities with highest positive sentiment scores} \\ \hline
Dusty Springfield & 0.75 & 40 & 4 \\ 
La Masia & 0.72 & 154 & 4 \\ 
Michael Polish & 0.72 & 58 & 4 \\ 
Federico Fernandez & 0.67 & 11 & 4 \\
Banqueting House & 0.67 & 90 & 4 \\ \hline
\multicolumn{4}{|c|}{Entities with highest negative sentiment scores} \\ \hline
New Jersey State Police	&-0.8&	7	&4\\
Vladikavkaz	&-0.73& 33&	4\\
Bureau of Consumer Protection&	-0.64&	568&	4\\
Georgia Diagnostic and Classification State Prison&	-0.63&	1,528	&4\\
Broadstairs&	-0.63	&3,174&	6\\\hline
\end{tabular}
\end{table}

It is worth noting that many of the most commented entities have an overall neutral or only very slightly polarized sentiment among \emph{Daily Mail} readership. And vice versa, Web sites where entities are mentioned the most usually keep a neutral or at most slightly positive or negative sentiment. For example, on Fig.~\ref{fig:platform} the largest circles tend to be in the middle. This could be due to the averaging effect over time: Web sites that mention a given entity a lot publish news with different sentiment, about positive and negative events, so the sentiment cancels out over time. This also implies that entities mentioned in fewer articles will be more polarized since the sentiment will be more likely due to only a few newsworthy events; results shown in Table~\ref{tab:sententities} indeed confirm this hypothesis. 
This leads to the idea of time series sentiment analysis that could be performed by combining NLP and time series analysis techniques.

One important task for a media outlet is to find and analyze its \emph{influencers}; an entire field of influencer marketing focuses on influential people rather than on target markets. For this purpose, we compare different measures of influence: total number of comments, number of replies, combined number of likes for all comments, number of dislikes, and three adaptations of the Hirsch index (\emph{h-index})~\cite{hirsch2005index}, a well-known bibliometric score designed to characterize the scientific output of a researcher by jointly measuring the author's productivity (number of papers) and impact of the author's work (number of citations). Following \citet{grvcar2017stance}, we define a user with an index of $h$ as a user that has posted $h$ comments each of which has received a given interaction mark (reply, like, or dislike) at least $h$ times.

\begin{figure}[t]
 	\centering    
      \includegraphics[width=0.9\linewidth]{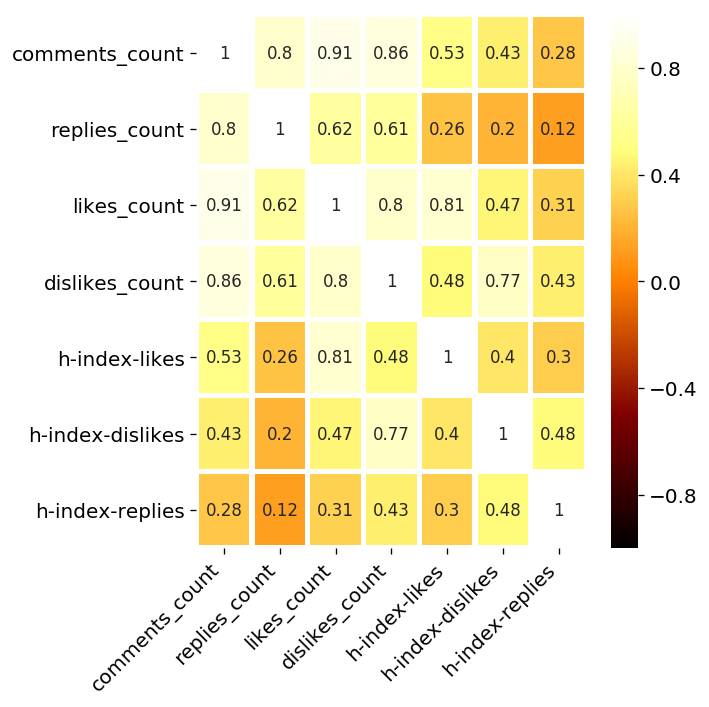}
\captionof{figure}{Correlation Matrix.}
\label{fig:corr}
\setlength{\tabcolsep}{3pt}
\captionof{table}{Stats for all and top 500 Daily Mail commenters.}
\label{tab:users}
\begin{tabular}{|l|c|c|}
\hline
 & \textbf{All} & \textbf{Top 500} \\ \hline
\# users & 54,024 & 500\\  
\# comments & 2,150,178 & 281,019 \\ 
\# replies & 1,563,267 & 233,805 \\ 
\# likes & 145,800,587 & 17,147,338 \\ 
\# dislikes & 37,550,303 & 4,891,883 \\ 
\hline
mean \# comments & 40 & 562 \\ 
mean \# replies & 29 & 468 \\ 
mean \# likes & 2,699 & 34,295 \\ 
mean \# dislikes & 695 & 9,784 \\
mean \texttt{h-index-likes} & 19.1 & 66.5 \\
mean \texttt{h-index-dislikes} & 9.7 & 35.9 \\
mean \texttt{h-index-replies}& 1.5 & 2.4 \\ 
\hline
\end{tabular}
\end{figure}

So which measures are most suitable for identifying influencers? We considered top 500 users with respect to the number of comments written. Table~\ref{tab:users} presents the statistics for this set of users and, for comparison, for the entire dataset. We note that the names and locations of users are gathered manually for this table using the publicly available data from the \emph{Daily Mail} user profiles.
Top~500 users wrote 13.1\% of all comments and received 11.8\% and 13\% of likes and dislikes, respectively. To identify similarities between measures of influence, Fig.~\ref{fig:corr} reports Pearson correlation coefficients between them: 
\begin{inparaenum}[(1)]
\item \texttt{comments\_count} has, naturally, a high correlation ($>$0.80) with \texttt{replies\_count}, \texttt{likes\_count}, and \texttt{dislikes\_count};
\item correlations between \texttt{h-index-replies} and all other measures do not exceed $0.6$, also expected since comments get less than $3$ replies on average;
\item there is a high correlation ($0.81$) between \texttt{h-index-likes} and \texttt{likes\_count}, while correlation between \texttt{h-index-likes} and \texttt{comments\_count} is only $0.53$, and a similar effect also holds for dislikes.
\end{inparaenum}
In general, our experiments show that measures based on likes and dislikes are substantially different from measures based on counting comments or replies.




Moreover, different measures can be indicative of real influence in different domains; e.g., in politics people want to state their opinions and thus write comments, while in celebrity news a like usually suffices. To see that, Table~\ref{tab:exusers} shows some top users ranked by four different metrics. 
Top users ranked by \texttt{comments\_count} and \texttt{replies\_count} discuss articles about politicians and international news (the most discussed entities in Table~\ref{tab:entities} are also mostly political). Top users ranked by \texttt{h-index-likes} are discussing popular celebrities. The case of \texttt{h-index-dislikes} is of separate interest because we see the most controversial topics rising to the top, such as Northern Ireland issues or supporting sports teams, where the body of commenters is naturally divided.
Note, however, that the top Daily Mail commenters are typically not the most influential by h-index metrics.

\begin{table*}[t]\centering
\setlength{\tabcolsep}{2pt}
\caption{The most influential Daily Mail users.}
\label{tab:exusers}
\scriptsize
\begin{tabular}{|p{2.7cm}|cccccc|p{7.3cm}|}
\hline
\textbf{User, City} & \multicolumn{6}{|c|}{\textbf{Number of}} &  \textbf{Most discussed entities} \\
& \textbf{comm.} & \textbf{repl.} & \textbf{likes} & \textbf{h-lk.} & \textbf{dislikes} & \textbf{h-dis.} & \\\hline
\multicolumn{8}{|c|}{Users ranked by \texttt{comments\_count}} \\ \hline
Dave, Gosport & 3,475 & 2,932 & 203,519 & 136 & 62,931 & 72 & Donald Trump, United States, United Kingdom, Russia, Paris\\
David, Dunmow & 2,838 & 1,829 & 162,833 & 132 & 46,068 & 69 & London, Donald Trump, United States, Chelsea, United Kingdom \\
John, Hong Kong & 2,296 & 1,633 & 177,209 & 140 & 36,524 & 62 & Donald Trump, United States, London, United Kingdom, North Korea\\
\hline
\multicolumn{8}{|c|}{Users ranked by \texttt{replies\_count}} \\ \hline
Dave, Gosport & 3,475 & 2,932 & 203,519 & 136 & 62,931 & 72 & Donald Trump, United States, United Kingdom, Russia, Paris\\
Type O Neg, Cheshire	&1,394 & 2,221	&62,470 & 90 	&14,489 & 42 & London, Catherine  Middleton, Martin McGuinness, France, Paris \\
Adam March, Kingston & 1,513 & 2,185 &	62,085 & 85 & 	14,916 & 36 & London, Manchester, United Kingdom, Instagram, Catherine  Middleton \\
\hline
\multicolumn{8}{|c|}{Users ranked by \texttt{h-index-likes}} \\ \hline
Bighoss, London&	729 & 20 &	129,292 & 175 &	13,908 & 46 & Kendall Jenner, Instagram, Kourtney Kardashian, Barcelona\\ 
Paul, Lansdale&  	1,986 & 942 &	182,944 & 157 &27,060 & 55 & Kim Kardashian, Donald Trump, North Korea, London, Britain\\
JennyO82, NYC& 	632 & 29 &	95,950 & 151 &	16,611 & 56 & Instagram, Ivanka Trump, Los Angeles, Bella Hadid, Oscar, Cannes\\
 \hline
\multicolumn{8}{|c|}{Users ranked by \texttt{h-index-dislikes}} \\
\hline
JayR$\_$, Monaco &    433 & 74  & 20,341 & 60 &50,364 & 99 & Barcelona, Dortmund, Madrid, Cristiano Ronaldo, Liverpool, NHS \\ 
LordBrendan, England& 167 & 15 &	24,314 & 56	&40,264 & 94	& Martin McGuinness, IRA, Barack Obama, Northern Ireland\\ 
LucidLucinda, Kensington & 353 & 20	&101,978 & 29 & 24,796 & 78 & London, Donald Trump, Manchester, Jeremy Corbyn, Brexit \\
\hline
\end{tabular}

\end{table*}

\subsection{Sentiment Evaluation of Brexit}\label{sec:brexit}

\begin{figure}[t]\centering
  	\includegraphics[width=0.87\linewidth]{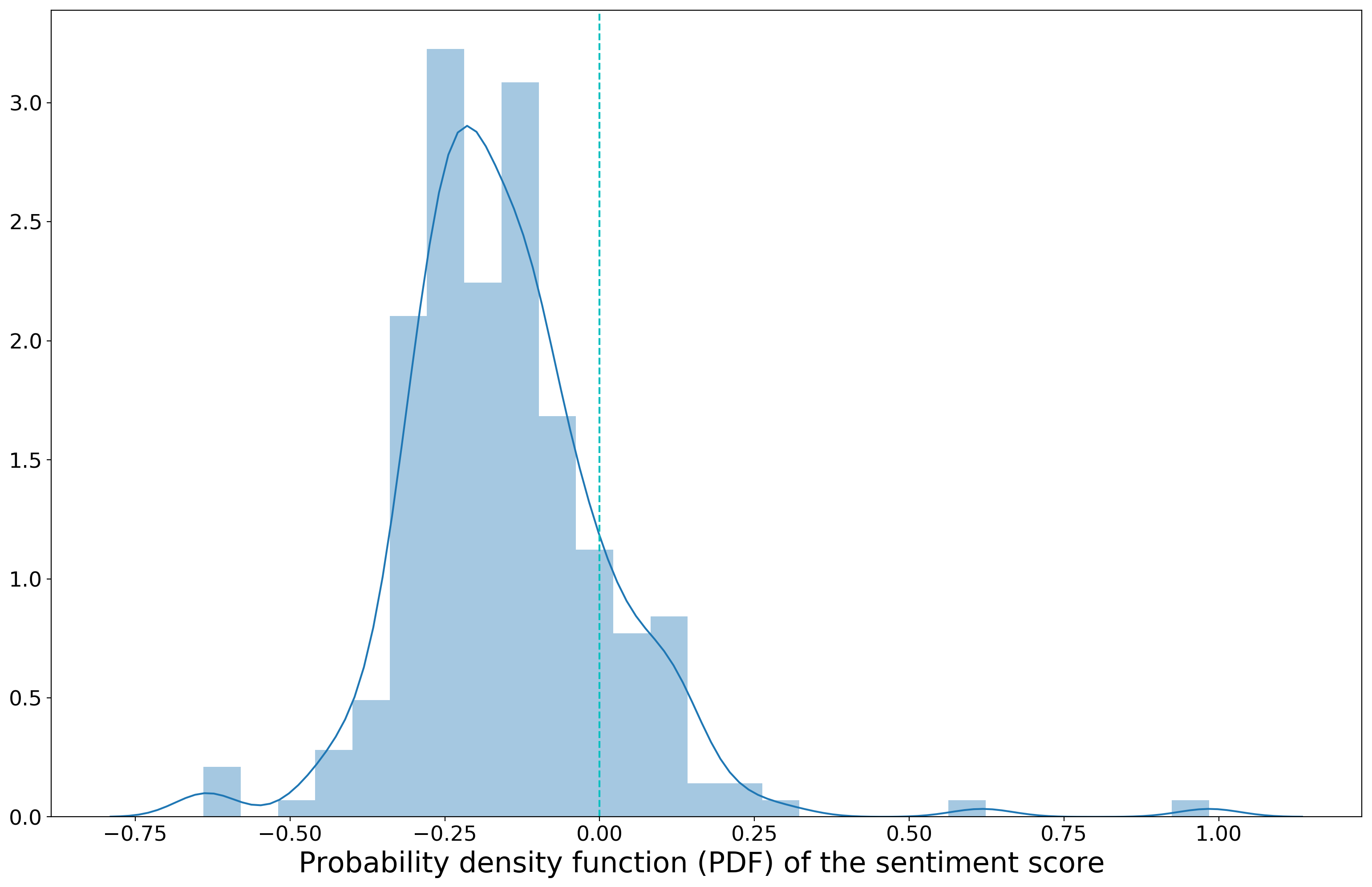}
	\caption{Probability density function (PDF) of the sentiment score.}
	\label{fig:brexit-pdf}
	\includegraphics[width=0.88\linewidth]{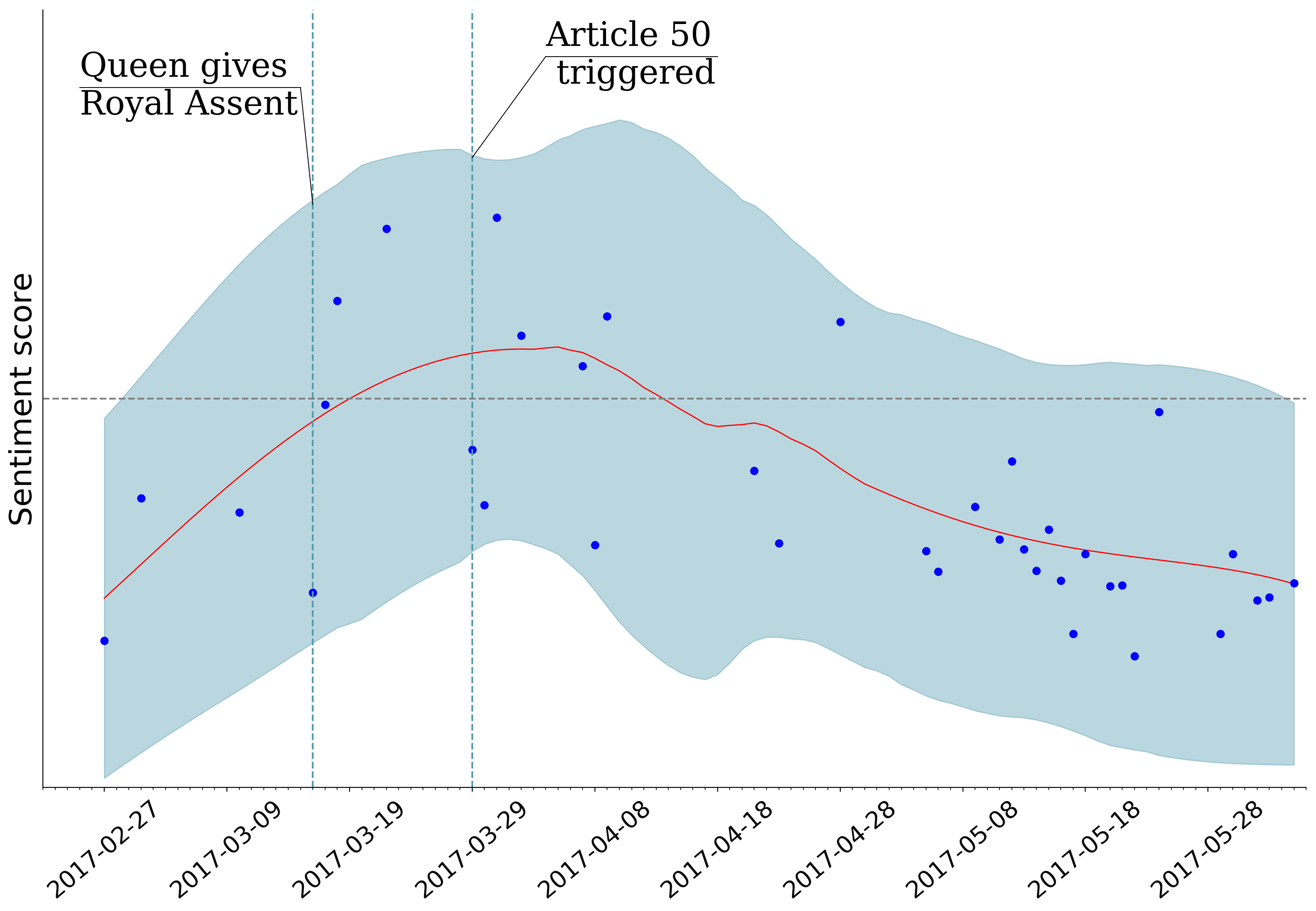}
	\caption{The sentiment evaluation of Brexit over time.}
	\label{fig:brexit}
\end{figure}

For this case study, we evaluated $87{,}904$ user comments linked to $200$ articles about Brexit, the British referendum to leave the EU, with $439.52$ comments per article on average. We computed the average sentiment of comments for each article. 
Cumulative sentiment distributions
of \emph{Daily Mail} users regarding Brexit 
are presented on 
Figs.~\ref{fig:brexit-pdf} and \ref{fig:brexit} show the results. 
Fig.~\ref{fig:brexit} shows sentiment scores for each day as average sentiment score across all articles published on this day (blue dots). To reduce random fluctuations and see sentiment trends through time, we applied linear interpolation to augment the number of available data points for smoothing and a smoothing filter by~\citet{savitzky1964smoothing}. The smoothed sentiment score is shown with the red line; we also show standard deviations for the smoothed line (light blue area).
%
%

Fig.~\ref{fig:brexit} shows the probability density function of the sentiment distribution of posts, with scores from $-1$ (very negative) to $1$ (very positive), with a slightly negative overall pattern, and 
sentiment changes over time. 
On March 28, 2017, Theresa May signed a letter invoking Article 50 that formally began the UK's departure from EU. At that day, overall sentiment was neutral. Starting from April 7, 2017, sentiment score tended to decrease from neutral ($\approx$0.05) to slightly negative ($\approx$-0.2). 
%
We have also analyzed opinion polls on whether the UK was right to decide to leave the EU conducted by \emph{YouGov}. 
Starting from April 26, 2017, it appears that more people thought the {Brexit} decision was wrong~\cite{YGPolls:2018,Curtis:2017}, so sentiment changes found for \emph{Daily Mail} comments do correlate with \emph{YouGov} opinion polls. It would be interesting to undertake a larger study over a period of several years; previous studies analyzed relations between the mood on \emph{Twitter} and the referendum outcome~\cite{agarwal2018geospatial,grvcar2017stance}, but the tweets they used date only from May-June 2016.

\subsection{Sentiment Evaluation of an Influencer}

In this case study, we investigate how negative comments are influenced by scandals and, generally, press surrounding the entity. We have investigated the case of Kendall Jenner, one of 2017's most popular Instagram influencers and thus a bankable bet for brands and marketers looking to leverage an influencer's following for engagement. 
We expanded the list of websites for this case study 
to collect opinions not only from Britain; in total, we found $115{,}582$ user comments linked to $1{,}009$ articles from $129$ websites, dating from February 20 to May 25, 2017, with $115$ comments per article on average. 

\begin{figure}[t]\centering
  	\includegraphics[width=0.87\linewidth]{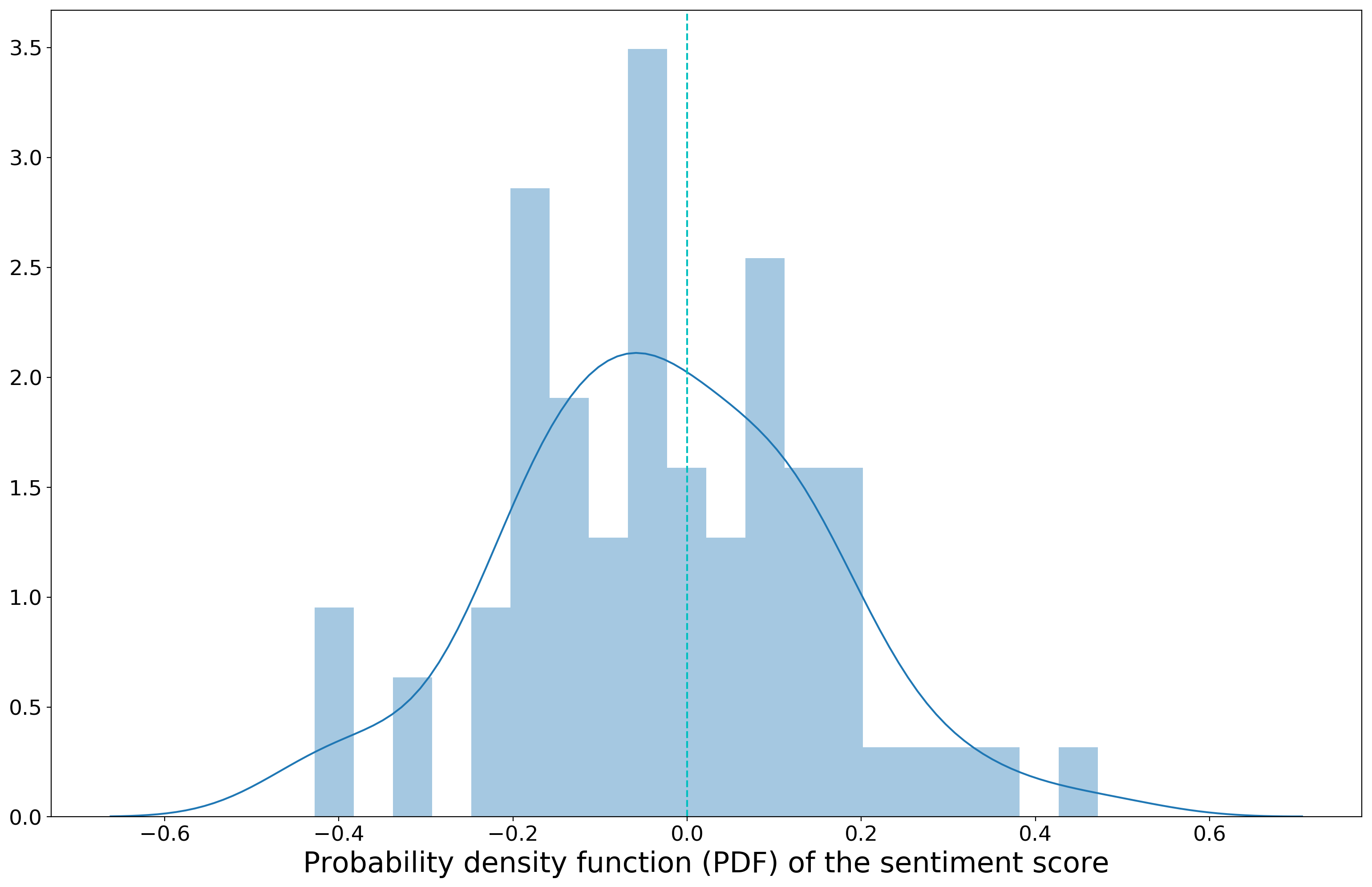}
    \caption{Probability density function (PDF) of the sentiment score.}
    \label{fig:kendall-pdf}
	\includegraphics[width=0.88\linewidth]{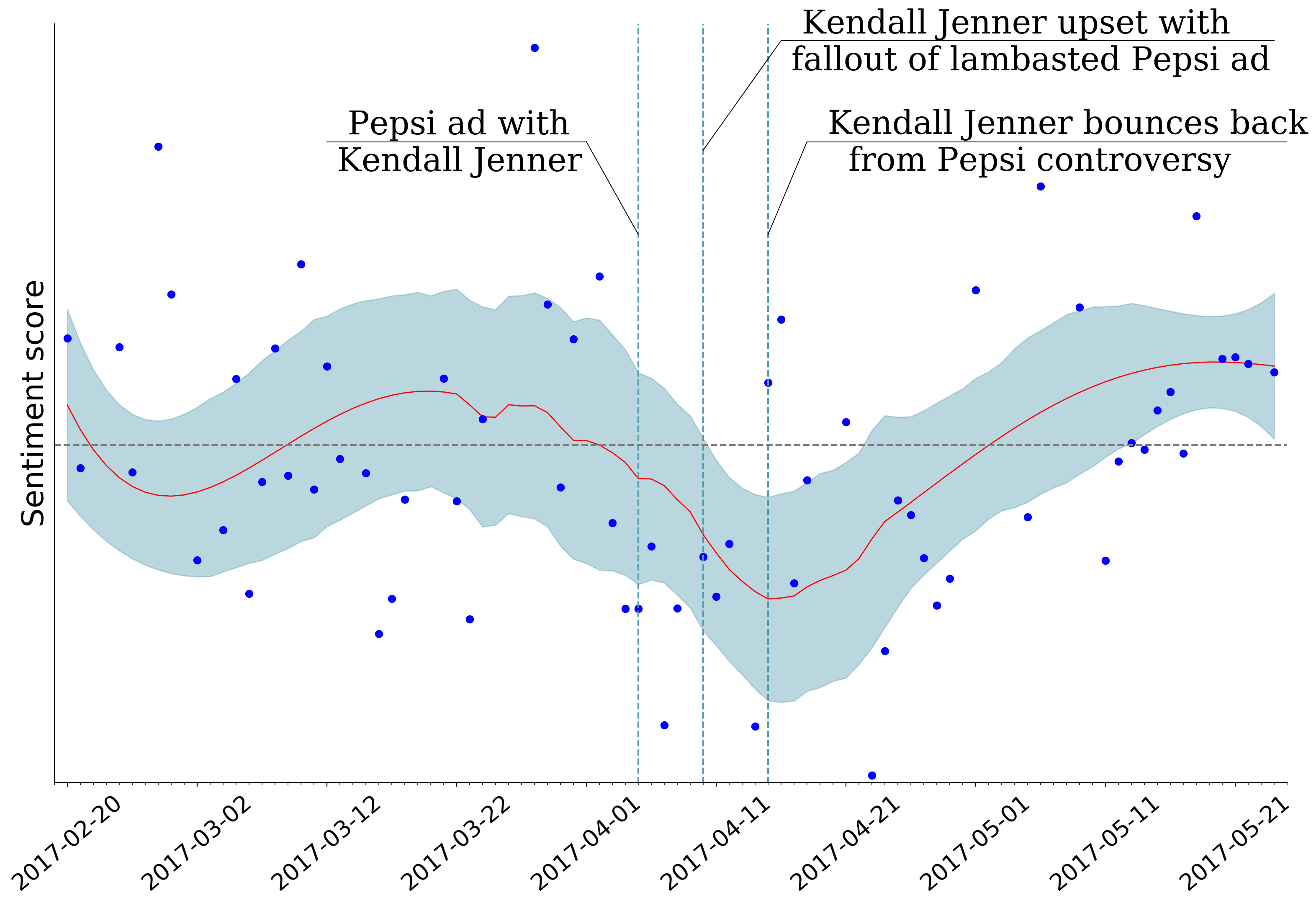}
	\caption{Sentiment evaluation of Kendall Jenner over time.}
	\label{fig:kendall}
\end{figure}

Fig.~\ref{fig:kendall-pdf} and~\ref{fig:kendall} show the sentiment distribution of posts and its changes over time. There is a clear shift starting in the first week of April: on April 5, \emph{Pepsi} released a commercial featuring Kendall Jenner in a multiracial protest. After massive reaction from groups such as \emph{Black Lives Matter}, \emph{Pepsi} apologized and pulled the advert less than 24 hours after its release. For several weeks, Jenner was a target of ridicule, and comments' sentiment decreased to negative ($\approx$-0.18); e.g., a \emph{Daily Mail} comment ``Clueless stupid girl makes clueless stupid commercial for clueless stupid company. Way to go, Pepsi'' gathered $2826$ likes and only $111$ dislikes. 
By the end of the month the effect waned, and sentiment became neutral again. 

\section{Conclusion}
\label{sec:concl}

In this work, we have presented the \emph{CommentsRadar} engine that collects articles and user comments from large Web sites, analyzes them with state of the art NLP models, and allows to draw important conclusions. With qualitative experiments shown in Section~\ref{sec:eval}, we have confirmed that the \emph{CommentsRadar} approach based on state of the art tools for named entity recognition, named entity linking, and sentiment analysis is already a suitable tool for discovering influencers in media outlets and analyzing sentiment over time for entities that appear in the news. 
Future directions for research include representing entities by both names and relations in category hierarchies, filtering offensive comments, detecting and tracking events over time, and clustering articles according to events.


%
\bibliographystyle{ACM-Reference-Format}
\bibliography{ml}

\end{document}